\documentclass[english,aps,superscriptaddress]{revtex4}
\usepackage[T1]{fontenc}
\usepackage[latin9]{inputenc}
\usepackage{babel}
\usepackage{float}
\usepackage{amsmath}
\usepackage{graphicx}
\usepackage[unicode=true]
 {hyperref}

\makeatletter
\@ifundefined{textcolor}{}
{%
 \definecolor{BLACK}{gray}{0}
 \definecolor{WHITE}{gray}{1}
 \definecolor{RED}{rgb}{1,0,0}
 \definecolor{GREEN}{rgb}{0,1,0}
 \definecolor{BLUE}{rgb}{0,0,1}
 \definecolor{CYAN}{cmyk}{1,0,0,0}
 \definecolor{MAGENTA}{cmyk}{0,1,0,0}
 \definecolor{YELLOW}{cmyk}{0,0,1,0}
}

\makeatother

\begin{document}

\title{Absorption and eigenmode calculation for one-dimensional periodic
metallic structures using the hydrodynamic approximation}

\author{Avner Yanai}

\affiliation{Department of Applied Physics, The Benin School of Engineering and
Computer Science, The Hebrew University of Jerusalem, Israel}

\author{N. Asger Mortensen}

\affiliation{DTU Fotonik, Department of Photonics Engineering, Technical University
of Denmark, DK-2800 Kongens Lyngby, Denmark}

\author{Uriel Levy}

\email{ulevy@mail.huji.ac.il}

\selectlanguage{english}%

\affiliation{Department of Applied Physics, The Benin School of Engineering and
Computer Science, The Hebrew University of Jerusalem, Israel}
\begin{abstract}
We develop a modal method that solves Maxwell's equations in the presence
of the linearized hydrodynamic correction. Using this approach, it
is now possible to calculate the full diffraction for structures with
period of the order of the plasma wavelength, including not only the
transverse but also the longitudinal modes appearing above the plasma
frequency. As an example for using this method we solve the diffraction
of a plane wave near the plasma frequency from a bi-metallic layer,
modeled as a continuous variation of the plasma frequency. We observe
absorption oscillations around the plasma frequency. The lower frequency
absorption peaks and dips correspond to lowest longitudinal modes
concentrated in the lower plasma frequency region. As the frequency
is increased, higher order longitudinal modes are excited and extent
to the region of higher plasma frequency. Moreover, examination of
the propagation constants of these modes reveals that the absorption
peaks and dips are directly related to the direction of phase propagation
of the longitudinal modes. Furthermore, we formulate a variant of
the Plane Wave Expansion method, and use it to calculate the dispersion
diagram of such longitudinal modes in a periodically modulated plasma
frequency layer.
\end{abstract}
\maketitle

\section{introduction}

Along with advances in nano-plasmonics, plasmonic devices reach length
scales for which non-local effects of the metal electric permittivity
function may no longer be neglected. For noble metals with critical
dimensions in the sub-10 nm regime, the longitudinal plasmonic response
exhibits spatial dispersion. This deviation from the ordinary local
approximation, requires modification of known analytical and numerical
tools. The hydrodynamic non-local model \cite{forstmann-book,boardman-book,pitarke,feibelman}
can be regarded as a simple approach (compared to more complex, quantum
models). However, it successfully reproduces experimental results
obtained for thin layered metals \cite{Anderegg} and offers a qualitative
explanation for the blue shifting of the localized surface-plasmon
resonance observed in silver nanoparticles \cite{ciraci,raza2}. These
results can not be explained with local models. In this paper, we
study the response of a metallic layer with periodic variation of
the free-carrier density, under the hydrodynamic approximation. While
the hydrodynamic model fails to account for quantum-size effects,
such as quantum tunneling \cite{stella-hydro-quantum,teperik-quantum-corrected,esteban-quantum-corrected},
it is a well established model for the dimensions studied here. Until
now, various numerical algorithms that solve Maxwell's equations with
the hydrodynamic correction have been reported \cite{toscano,ruppin-1,hiremath,yannopapas,mochan-tmm,deabajo,fernandez-dominguez,raza,toscano-2}.
In this paper, we provide a rigorous numerical approach, that allows
the calculation of 1D periodic structures. Our method relies on the
Fourier Modal Method (FMM) also known as the Rigorous Coupled Wave
Analysis (RCWA) method \cite{moharam,lalanne,lifengli2}. This method
can be regarded as semi-analytic in the sense that not only the field
distribution is calculated, but also the propagation constants and
the eigenmodes of the periodic structure are obtained, allowing to
derive additional physical insight (see e.g. \cite{lalanne-2}). Adding
the hydrodynamic terms to the ordinary FMM formulation, allows us
to utilize some of the strengths that are offered by FMM. The paper
is structured as follows. In Section \ref{sec:fmm-formulation} the
FMM with the additional hydrodynamic terms is presented. In addition,
we formulate the band diagram dispersion calculation of the longitudinal
modes. In Section \ref{sec:results}, results based on this framework
are shown. Section \ref{sec:conclusions} concludes the paper.

\section{FMM with the hydrodynamic correction\label{sec:fmm-formulation}}

\begin{figure}
\begin{centering}
\includegraphics[scale=0.3]{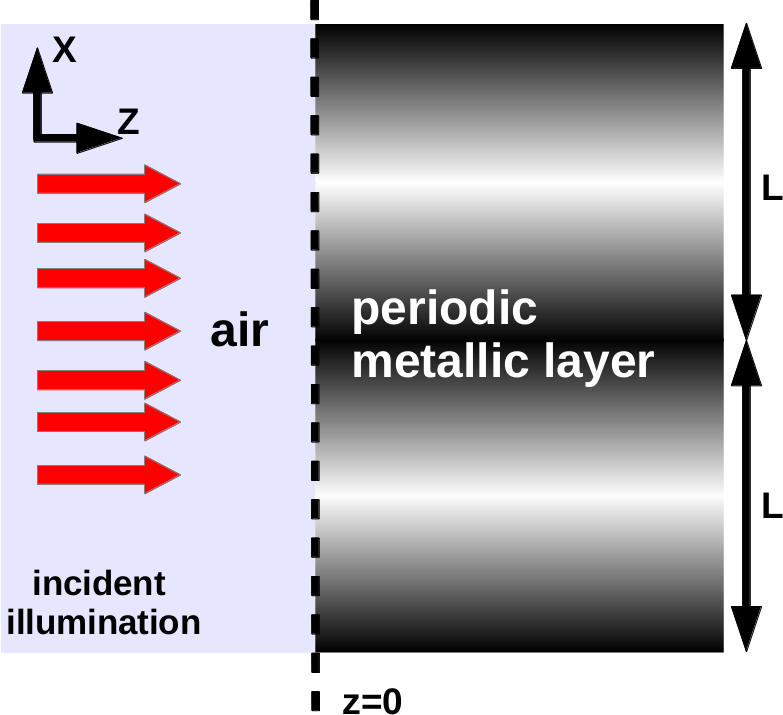}
\par\end{centering}

\caption{Schematic showing two unit cells of the diffraction problem geometry,
with light incident from left to right. The air/metal interface is
located at z=0, and the direction of periodicity is along the vertical
(i.e. x) axis. The region with reduced free-carrier density is in
the center of the unit cell (bright), whereas the boundaries of the
unit cell are the region with higher free-carrier density (dark).\label{fig:geometry}}
 
\end{figure}
First, we briefly review the essential basics of the FMM. Further
details can be found in several references, e.g. \cite{moharam,lalanne}.
In its most common formulation, the FMM uses a Floquet-Bloch expansion
within a unit cell $0\le x<L$ (see schematic in Fig. \ref{fig:geometry}),
to represent Maxwell's equations in each z-invariant periodic layer.
Afterwards, the eigenmodes and eigenvalues of the fields are calculated
by solving an eigenvalue equation. We now elaborate on these principles.
The Floquet-Bloch condition implies that the wavevectors in the $x$
direction are given by $k_{x,m}=k_{x,0}+mK$, where $K=2\pi/L$ is
the grating vector and $k_{x,0}$ is the \textquotedblleft{}zero-order\textquotedblright{}
term. With the FMM, Maxwell's equations are solved for each locally
z-invariant layer, from an eigenvalue equation of the form $\frac{\partial}{\partial z}\mathbf{F}=\mathbf{A}\,\mathbf{F}$.
Here, $\mathbf{F}$ is a column vector of the Fourier components of
the fields and $\mathbf{A}$ is an operator matrix defined by Maxwell's
equations. The propagation constants $k_{z,n}$, are obtained by solving
this eigenvalue equation. In order to solve the eigenvalue equation
numerically, one must truncate the number of Fourier components to
some finite number of $N$ elements, with $-\left\lfloor N/2\right\rfloor \leq m\leq\left\lfloor N/2\right\rfloor $.
Generally, the solution converges to the exact solution by increasing
$N$. In order to solve a diffraction problem, the fields in adjacent
layers are matched by employing the proper boundary conditions. By
this matching procedure, a mode amplitude constant $C_{n}$ is solved
for the $n^{th}$ eigenmode \cite{moharam}. In the following subsection
we present the derivation of the matrix operator $\mathbf{A}$ in
the presence of the hydrodynamic correction.

\subsection{Maxwell's equations with the hydrodynamic correction}

In each z-invariant layer, and for a single frequency component $\omega$,
Maxwell's equations with the linearized hydrodynamic correction are
given by \cite{boardman-ruppin,hiremath}: 

\begin{subequations} \label{one-all-eqn}

\begin{eqnarray}
\nabla\times E-j\omega\mu_{0}H & = & 0\label{eq:1}
\end{eqnarray}

\begin{eqnarray}
\nabla\times H+j\omega\varepsilon_{0}E+eN_{0}(x)v & = & 0\label{eq:2}
\end{eqnarray}

\begin{eqnarray}
\varepsilon_{0}\nabla\cdot E+en_{1} & = & 0\label{eq:3}
\end{eqnarray}

\begin{eqnarray}
\nabla\cdot\left[N_{0}(x)v\right]-j\omega n_{1} & = & 0\label{eq:4}
\end{eqnarray}
\begin{eqnarray}
-jm\omega(N_{0}(x)v)+m\gamma(N_{0}(x)v)+N_{0}(x)eE+m\beta^{2}\nabla n_{1} & = & 0\label{eq:5}
\end{eqnarray}
\end{subequations}where $N_{0}(x)$ is a periodic function of the
density of free electrons in equilibrium, $n_{1}$ is the first-order
non-equilibrium correction to the equilibrium electron density and
likewise $v$ is the first-order non-equilibrium electron velocity
while there are no equilibrium currents. Furthermore, the strength
of the non-local response is governed by $\beta^{2}=\frac{3}{5}v_{F}^{2}$.
The electron mass is denoted by $m$. The case that $N_{0}$ varies
with $x$ while $\beta$ is constant, can be regarded as a toy model
for the scenario in which two metals with different plasma frequencies
fill the unit cell, with continuous variation of the free carrier
density. We solve the set of Eq. \eqref{one-all-eqn} for TM polarization
{[}i.e. $H(x,y,z)=\hat{y}H_{y}(x,z)$ and $E(x,y,z)=\hat{x}E_{x}(x,z)+\hat{z}E_{z}(x,z)${]},
as only this polarization supports longitudinal modes. As explained
above, in order to solve Eq. \eqref{one-all-eqn} with standard FMM
formulation, we need to isolate all $\frac{\partial}{\partial z}$
dependencies to obtain an eigenvalue equation. For convenience, we
introduce $\hat{\omega}^{2}=\omega(\omega+i\gamma),\:\tilde{\beta}^{2}=\beta^{2}/c^{2},\tilde{\:\omega}_{p}^{2}=\omega_{p}^{2}/c^{2}=\frac{N_{0}e^{2}}{m\varepsilon_{0}}/c^{2},\:\hat{k}_{0}=\hat{\omega}/c,\hspace{1em}\tilde{\mu}_{0}=c\mu_{0}\text{, and }\tilde{\varepsilon}_{0}=c\varepsilon_{0}=\tilde{\mu}_{0}^{-1}.$
Furthermore, we define the hydrodynamic current as $J=eN_{0}v$ and
$J(x,y,z)=\hat{x}J_{x}(x,z)+\hat{z}J_{z}(x,z)$. Since we have spatial
harmonic variations, we straightforwardly make the following substitutions
for the derivatives: $\partial/\partial_{z}=jk_{0}k_{z}$ and $\partial/\partial_{x}=jk_{0}k_{x}$.
Making these substitutions and performing algebraic manipulations
described in some detail in Appendix \ref{sec:eigenvalue-derivation},
we arrive at the eigenvalue equation in matrix form:

\begin{multline}
\left[\begin{array}{c}
\mathbf{E_{x}}\\
\mathbf{\nabla\cdot J}
\end{array}\right]\left[\mathbf{K_{z}^{2}}\right]=\left[\begin{array}{cc}
\tilde{\mu}_{0}\mathbf{I} & \mathbf{K_{x}}\\
\hat{k}_{0}^{2}\tilde{\beta}^{-2}\mathbf{K_{x}} & \tilde{\varepsilon}_{0}\tilde{\beta}^{-2}\left(\hat{k}_{0}^{2}\mathbf{I}-\mathbf{\Omega_{p}^{2}}\right)
\end{array}\right]\left[\begin{array}{cc}
\tilde{\varepsilon}_{0}\hat{k}_{0}^{-2}\left(\hat{k}_{0}^{2}\mathbf{I}-\mathbf{\Omega_{p}^{2}}\right) & -\hat{k}_{0}^{-2}\tilde{\beta}^{2}\mathbf{K_{x}}\\
-\mathbf{K_{x}} & \tilde{\mu}_{0}k_{0}^{-2}\mathbf{I}
\end{array}\right]\left[\begin{array}{c}
\mathbf{E_{x}}\\
\mathbf{\nabla\cdot J}
\end{array}\right]\label{eq:fmm-final}
\end{multline}
Here, $\mathbf{E_{x}}$ and $\mathbf{\nabla\cdot J}$ are the eigenvector
matrices of \textbf{$E_{x}$} and $\nabla\cdot J$ respectively. $\mathbf{K_{x}}$
and $\mathbf{K_{z}}$ are diagonal matrices with elements $k_{x,m}$
and $k_{z,n}$ respectively, and the identity matrix is $\mathbf{I}$.
$\mathbf{\Omega_{p}^{2}}$ is the Toeplitz matrix with elements corresponding
to the Fourier components of $\tilde{\omega}_{p}^{2}(x)$. The matrices
$\mathbf{I}$ , $\mathbf{\Omega_{p}^{2}}$ , $\mathbf{K_{x}}$, $\mathbf{E_{x}}$
and \textbf{$\mathbf{\nabla\cdot J}$} are of size $N\times N$ ,
while $\mathbf{K_{z}}$ is a $2N\times2N$ matrix and the overall
number of eigenmodes obtained from Eq. \eqref{eq:fmm-final} is $2N$.
However, in local media, the number of eigenmodes is equal to the
truncation number i.e. $N$ \cite{moharam}. Therefore, a total of
$3N$ mode amplitude constants need to be found when matching the
fields at the interface of a local layer with a layer with non-local
response. The ordinary boundary conditions demanding continuity of
the tangential field components $E_{x}$ and $H_{y}$ provide only
$2N$ equations. To match the ``missing'' $N$ amplitude constants,
an additional boundary condition (ABC) is required. This is very similar
to the known case of matching the field amplitudes between two homogeneous
local and non-local layers \cite{melnyk}. For simplicity, we consider
an air/metal interface. For this case, the boundary conditions are
the continuity of $E_{x}$, $J_{z}$ and $E_{z}$ across the interface
\cite{forstmann-book,yan-hyperbolic,moreau}. We note however, that
the ABC does not change in any way the eigenvalue equation of the
periodic metallic layer, which is a direct solution of Maxwell's equations
with the hydrodynamic correction. In Section \ref{sec:results-fmm},
we solve a full diffraction problem for the case of light incident
from air on a semi-infinite periodic non-local layer. At the interface
of both media, the mode amplitude constants are found by employing
these boundary conditions. In Appendix \ref{sec:s-matrix}, we present
a procedure based on the S-matrix algorithm \cite{lifengli2}, for
the calculation of the mode amplitudes. For completeness, this procedure
is formulated for the more general case of a non-local periodic layer
embedded in a local environment.

\subsection{Band diagram calculation \label{sec:pwm}}

In order to calculate the dispersion diagram of the longitudinal modes,
we follow an approach based on that in \cite{dispersive-band-diagram}.
This approach is a variant of the Plane Wave expansion Method (PWM)
\cite{photonics-crystal-book}. In contrast to the conventional PWM,
where the Bloch wavevector ($k_{x,0}$) is assumed and the frequency
($\omega$) is solved from an eigenvalue problem, in the revised PWM,
the frequency is assumed beforehand, and the phase difference of the
fields across the unit cell (known as the Bloch wavevector) is calculated.
We note that for nondispersive loss-less materials there is a freedom
to choose one or the other. However, for the dispersive metal it is
important that one solves for the complex Bloch wave vector while
treating the frequency as real (alternatively, one can also solve
for complex $\omega$, see Ref. \cite{fan}) . This is relevant for
the experimental situation where the structure is probed by a CW laser
with a well-defined frequency. We now show how the PWM variant can
be applied to calculate the dispersion diagram of a metallic structure
with the hydrodynamic correction. For simplicity, we assume a 1D case
with $k_{z}=0$ , and that $\omega_{p}$ is periodic with $x$. For
such a case, Eq. \eqref{one-all-eqn} (see also Eq. \eqref{revised-der-all})
reduces to three first order differential equations:\begin{subequations}
\label{pwm-all}
\begin{eqnarray}
k_{x}E_{z}+\tilde{\mu}_{0}H_{y} & = & 0\label{eq:band-1}
\end{eqnarray}

\begin{eqnarray}
\tilde{\varepsilon}_{0}\varepsilon_{T}E_{z}+k_{x}H_{y} & = & 0\label{eq:band-2}
\end{eqnarray}

\begin{equation}
E_{x}-\hat{k}_{0}^{-2}\tilde{\omega}_{p}^{2}(x)E_{x}-\hat{k}_{0}^{-2}k_{0}^{2}\tilde{\beta}^{2}k_{x}^{2}E_{x}=0\label{eq:band-3}
\end{equation}
\end{subequations}Here $\varepsilon_{T}\equiv1-\frac{\tilde{\omega}_{p}^{2}(x)}{\hat{k}_{0}^{2}}.$
Eq. \eqref{pwm-all} can be subdivided into two independent sets:
Eq. \eqref{eq:band-1} and \eqref{eq:band-2} describe the transverse
modes (no field components in the direction of the only non-zero k-vector
component, i.e. $k_{x})$ while Eq. \eqref{eq:band-3} defines the
dispersion law of the longitudinal modes. Moreover, by defining $\varepsilon_{L}\equiv1-\frac{\tilde{\omega}_{p}^{2}(x)}{\hat{k}_{0}^{2}-k_{0}^{2}\tilde{\beta}^{2}k_{x}^{2}}$,
Eq. \eqref{eq:band-3} can be re-written as $\varepsilon_{L}E_{x}=0$,
from which the familiar condition for longitudinal modes $\varepsilon_{L}=0$
is apparent. To solve Eq. \eqref{eq:band-3}, we define $k_{x}\equiv k_{x,0}+k_{x,m}$.
Here $k_{0}k_{x,0}L$ is the phase difference of the field $F$ between
the two boundaries of the unit cell according to: $F(x=L)=F(x=0)\exp(jk_{0}k_{x,0}L)$,
and $k_{x,m}=mK$. With these definitions, and the auxiliary field
quantity $\dot{E}_{x}\equiv\left[k_{x,0}+k_{x,m}\right]E_{x}$, we
split Eq. \eqref{eq:band-3} into two first order equations\begin{subequations}:\label{pwm-all-2}

\begin{equation}
\tilde{\beta}^{-2}k_{0}^{-2}\left(\hat{k}_{0}^{2}-\tilde{\omega}_{p}^{2}(x)\right)E_{x}-\left[k_{x,0}+k_{x,m}\right]\dot{E}_{x}=0\label{eq:pwm-all-2-1}
\end{equation}

\begin{equation}
\dot{E}_{x}=\left[k_{x,0}+k_{x,m}\right]E_{x}\label{eq:pwm-2-2}
\end{equation}
\end{subequations} Eq. \eqref{pwm-all-2} can be recast to the matrix
form:

\begin{equation}
\left[\begin{array}{c}
\mathbf{\dot{E}_{x}}\\
\mathbf{E_{x}}
\end{array}\right]\left[\mathbf{K_{x,0}}\right]=\left[\begin{array}{cc}
-\mathbf{K_{x}} & \tilde{\beta}^{-2}k_{0}^{-2}\left(\hat{k}_{0}^{2}\mathbf{I}-\mathbf{\Omega_{p}^{2}}\right)\\
\mathbf{I} & -\mathbf{K_{x}}
\end{array}\right]\left[\begin{array}{c}
\mathbf{\dot{E}_{x}}\\
\mathbf{E_{x}}
\end{array}\right]\label{eq:band-diagram-final}
\end{equation}
Here $\mathbf{K_{x,0}}$ is a $2N\times2N$ diagonal matrix with elements
corresponding to the phase difference between the boundaries of the
unit cell of each eigenmode, and $\mathbf{K_{x}}$ is an $N\times N$
diagonal matrix with elements $k_{x,m}=mK$. Eq. \eqref{eq:band-diagram-final}
can be identified as an eigenvalue equation from which the matrix
of phase constants $\mathbf{K}_{\mathbf{x,0}}$ can be obtained.

\section{Simulation results\label{sec:results}}

\subsection{Absorption spectrum of a semi-infinite metallic layer with sinusoidal
modulation of $\omega_{p}$\label{sec:results-fmm}}

\begin{figure}
\begin{centering}
\includegraphics[scale=0.7]{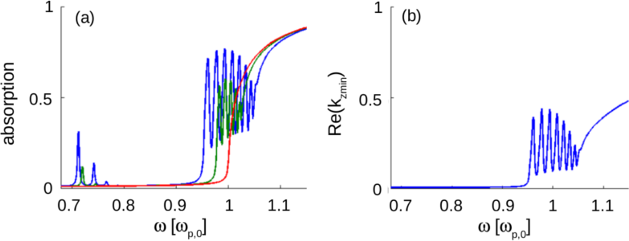}
\par\end{centering}

\caption{(a) Absorption spectrum of a semi-infinite metallic layer with $\lambda_{p}=\frac{2\pi c}{\omega_{p,0}}=5L$.
Blue line, $\omega_{p,1}^{2}=0.1\omega_{p,0}^{2}.$ Green line $\omega_{p,1}^{2}=0.05\omega_{p,0}^{2}$.
Red line: $\omega_{p,1}^{2}=0$\label{fig:Absorption-spectrum1} (b)
Real part of the propagation constant with absolute value closest
to zero, calculated for the case that $\omega_{p,1}^{2}=0.1\omega_{p,0}^{2}.$}
 
\end{figure}
\begin{figure}[H]
\begin{centering}
\includegraphics[scale=0.45]{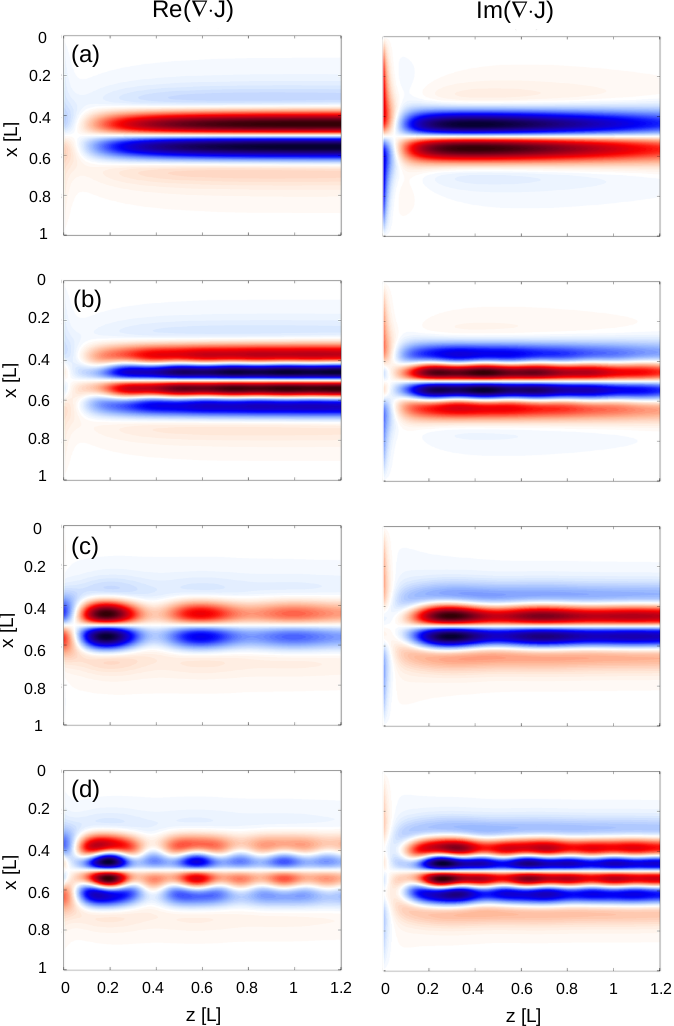}
\par\end{centering}

\caption{(a,b) Real and imaginary parts of $\nabla\cdot J$, calculated at
the absorption peaks, located at (a)\textbf{ }$\frac{\omega}{\omega_{p,0}}=\text{0.9605 and (b) \ensuremath{\frac{\omega}{\omega_{p,0}}}=0.9772 }$.
(c,d) Real and imaginary parts of $\nabla\cdot J$, calculated at
the absorption dips, located at (c)\textbf{ }$\frac{\omega}{\omega_{p,0}}=\text{0.9674 and (d) \ensuremath{\frac{\omega}{\omega_{p,0}}}=0.9841 }$
The air/metal interface is at z=0.\label{fig:Absorption-peaks-and-dips}}
\end{figure}

As first application of the modified FMM, we first consider the following
toy geometry: A TM plane wave is normally incident upon a semi-infinite
metallic layer, with modulation of the plasma frequency given by:
$\omega_{p}^{2}=\omega_{p,0}^{2}+\omega_{p,1}^{2}\cos(\frac{2\pi x}{L})$.
The material parameters are $\gamma=\omega_{p,0}/300$, $v_{F}=0.01c$
and $\lambda_{p}\equiv\frac{2\pi c}{\omega_{p,0}}=5L$. The calculation
has been repeated for the following three modulation amplitudes: $\omega_{p,1}^{2}=0.1\omega_{p,0}^{2},$
$0.05\omega_{p,0}^{2}$ and 0. The results are shown in Fig. \ref{fig:Absorption-spectrum1}(a).
Two sets of absorption peaks are observed: 1) Absorption peaks near
the frequencies $\sim\left(1/\sqrt{2}\right)\omega_{p,0}$ . These
are surface waves (surface plasmon polariton (SPP) like) that are
confined near the interface. 2) Absorption oscillations appearing
near $\omega_{p,0}$ , which are the consequence of longitudinal waves.
It is seen that the oscillation strength increases as the plasma frequency
modulation amplitude increases. In Fig. \ref{fig:Absorption-peaks-and-dips}(a,b)
we plot $\nabla\cdot J,$ for the two lowest frequency absorption
peaks ($\frac{\omega}{\omega_{p,0}}=\text{0.9605 and 0.9772})$, and
in Fig. \ref{fig:Absorption-peaks-and-dips}(c,d) we plot the same
quantities for the first two absorption dips ($\frac{\omega}{\omega_{p,0}}=\text{ 0.9674 and 0.9841})$
. Since $\nabla\cdot J$ is proportional to the induced charge density
(see Eq. \eqref{eq:4}), it can be seen that for the lower frequency
modes, the induced charge density concentrates in the middle of the
unit cell where the plasma frequency is minimal. This is consistent
with previous studies where a layer of metal with lower plasma frequency
was deposited on top of a higher plasma frequency metal. For such
a case, standing waves in the lower plasma frequency region, similar
to those in a 1D potential well were observed (see Ref. \cite{forstmann-book}
section 3.4). In addition, calculation of $H_{y}$ which is a transverse
field quantity (no magnetic field exists in a longitudinal mode),
shows that the magnetic field is negligible compared to $\nabla\cdot J$,
manifesting that the modes are almost completely longitudinal in nature.
To reveal the reason for the existence of the absorption peaks and
dips, we plot in Fig. \ref{fig:Absorption-spectrum1}(b) the real
part of the propagation constant with absolute value closest to zero
as a function of frequency, for the case that $\omega_{p,1}^{2}=0.1\omega_{p,0}^{2}.$
More mathematically stated, we define $k_{z,\text{min}}$ $\equiv\text{min}(|k_{z,n}|^{2})$
and plot Re$(k_{z,\text{min}})$. There is a clear correspondence
between the blue line in Fig. \ref{fig:Absorption-spectrum1}(a) and
Fig. \ref{fig:Absorption-spectrum1}(b). The dips and peaks of the
absorption spectrum are located at the minimas and the maximas of
Re$(k_{z,\text{min}})$ respectively. The reason for this is that
when Re$(k_{z})$ is relatively large, the longitudinal mode propagates
with significant phase accumulation along the $z$ axis and eventually
dissipates. On the other hand, when Re$(k_{z})\cong0$, the longitudinal
mode barely propagates into the metal, but rather has a standing wave
pattern along the $x$ axis (Supplemental Material \cite{supplemental-material}).
This analysis shows the strength of the FMM approach. Being a semi-analytical
method, it provides physical insight due to the calculation of modes
and propagation constants. 

\subsection{Absorption spectrum of an Au/Ag bi-metallic semi-infinite layer\label{sec:results-fmm-2}}

\begin{figure}
\includegraphics[scale=0.25]{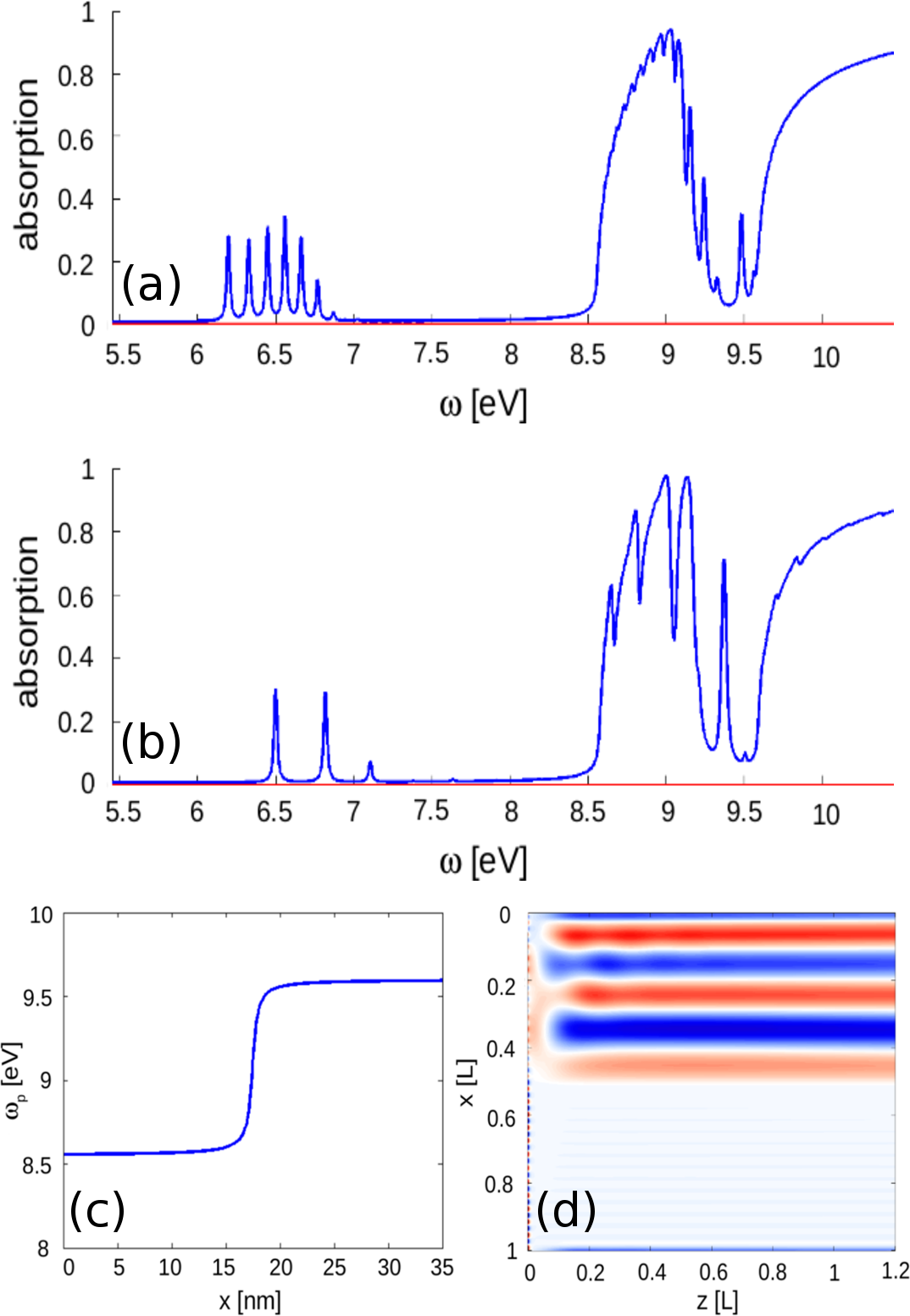}

\caption{(a) Absorption spectrum of an Ag/Au grating with $L=$35{[}nm{]} as
function of $\omega$. (b) Absorption spectrum of an Ag/Au grating
with $L=$10{[}nm{]} as function of $\omega$. (c) The distribution
of $\omega_{p}$ in the unit cell assumed for the calculation. (d)
The real part of $\nabla\cdot J$, calculated for $\omega=8.592${[}eV{]}
and $L=$35{[}nm{]}. \label{fig:step-function}}
\end{figure}

We now turn to analyze the case of an Au/Ag semi-infinite layer. We
assume $\omega_{p,\text{Au}}=8.55${[}eV{]}, $\omega_{p,\text{Ag}}=9.6${[}eV{]},
$\gamma_{\text{Au}}=\gamma_{\text{Ag}}=0.02${[}eV{]} and $v_{F,\text{Au}}=v_{F,\text{Ag}}=0.0047c$.
These parameters are from \cite{blaber}, with the simplifying assumption
that the damping in Au and Ag is the same. In the unit cell $0\le x<L$,
the plasma frequency is described by: $\omega_{p}^{2}(x)=\text{arctan}\left[(x-0.5L)(f/L)\right]\left(\omega_{p,\text{Au}}^{2}-\omega_{p,\text{Ag}}^{2}\right)/\pi+\left(\omega_{p,\text{Au}}^{2}+\omega_{p,\text{Ag}}^{2}\right)/2$.
This function results in continuous step-like profile. The advantage
of using such function is that it eliminates the need to take care
of the correct factorization rules of a piecewise discontinuous function
\cite{lifengli,lalanne}, and also provides a more realistic description
of the transition between the two metals. The parameter $f$ determines
the steepness of the transition between both media. In Fig. \ref{fig:step-function}(a)
we plot the absorption spectrum, for $f=100$ and $L=$35{[}nm{]}
. The step-like distribution of $\omega_{p}$ in the unit cell is
shown in Fig. \ref{fig:step-function}(c). Similarly to the case studied
in Section \ref{sec:results-fmm}, the absorption spectrum exhibts
peaks near $\left(1/\sqrt{2}\right)\omega_{p,\text{Au}}$ and $\left(1/\sqrt{2}\right)\omega_{p,\text{Ag}}.$
Additionaly, there are absorption peaks due to longitudinal modes
for $\omega_{p,\text{Au}}<\omega<\omega_{p,\text{Ag}}$. The reason
no absorption peaks are observed for $\omega>\omega_{p,\text{Ag}}$,
is that the unit cell is larger than the typical dimension ($\sim$10
{[}nm{]}) for which non-local effects are significant for these metals.
However, when $\omega_{p,\text{Au}}<\omega<\omega_{p,\text{Ag}}$,
longitudinal modes exist only in the Au layer, which for the assumed
unit cell size, is small enough to clearly observe longitudinal resonances.
Indeed, for a smaller unit cell with $L=10${[}nm{]}, resonances of
longitudinal modes for $\omega>\omega_{p,\text{Ag}}$ are observed
(see Fig. \ref{fig:step-function}(b)). In Fig. \ref{fig:step-function}(d),
we show $\nabla\cdot J$ calculated for $\omega=8.592$ {[}eV{]} and
$L=35${[}nm{]}. It can be observed that the longitudinal modes are
confined in the Au layer only. 

\subsection{Band diagram calculation}

We now turn into calculating the 1D dispersion diagram of longitudinal
modes, based on the formulation described in Section \ref{sec:pwm}.
We assume the following parameters: $\gamma=0$ (no losses), $v_{F}=0.01c$
and $\lambda_{p}=\frac{2\pi c}{\omega_{p,0}}=10L$.
\begin{figure}
\includegraphics[scale=0.5]{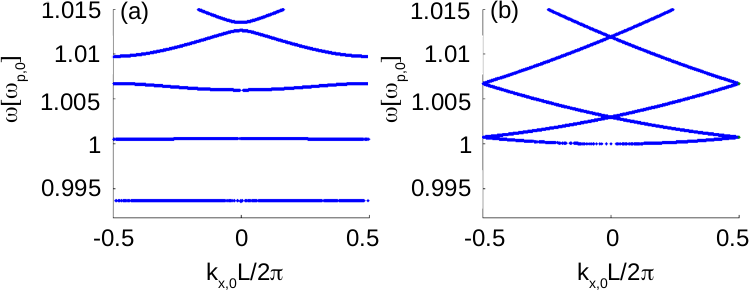}

\caption{Dispersion of the longitudinal modes. (a) periodic case with $\omega_{p,1}^{2}=0.02\omega_{p,0}^{2}$.
(b) uniform case $(\omega_{p,1}^{2}=0)$ \label{fig:band-diagram-dispersion}}
\end{figure}

In Fig. \ref{fig:band-diagram-dispersion}(a) we assumed $\omega_{p,1}^{2}=0.02\omega_{p,0}^{2}$.
For this case both the band gaps at the edges of the 1st Brillouin
Zone (BZ) and at $k_{x,0}=0$ are apparent. Moreover, the dispersion
of the lower order modes is flat, which is an indication for a very
low group velocity, regardless of the specific momentum value. This
is in contrast to the more conventional case of a periodic structure
which generates slow light only at the edges of the BZ. Losses (neglected
here for simplicity) however, cause broadening and enhance the group
velocity at the band edges \cite{slow-light-band} . In \ref{fig:band-diagram-dispersion}(b)
we assumed a uniform metallic medium having no modulation, i.e. $\omega_{p,1}^{2}=0$.
Obviously, for this case, no bandgaps are observed, as expected. 
\begin{figure}
	\includegraphics[scale=0.65]{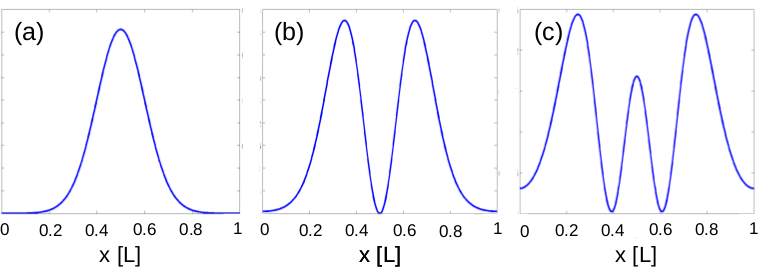}

\caption{The profile of $|J_{x}|^{2}$ for the 1st - 3rd bands of the dispersion
diagram plotted in Fig. \ref{fig:band-diagram-dispersion}(a)\label{fig:band-diagram-modes}.
(a) $\omega=0.9936\omega_{p,0}$ (b) $\omega=1.0005\omega_{p,0}$
(c) $\omega=1.0064\omega_{p,0}$ . All modes are calculated for $k_{x,0}=0.$}
\end{figure}

In Fig. \ref{fig:band-diagram-modes} we plot the mode profile of
$|J_{x}(x)|^{2}$ (which is proportional to the kinetic energy of
the charges) for the three first bands of the dispersion diagram plotted
in Fig. \ref{fig:band-diagram-dispersion}(a). As discussed before,
it is seen that the lowest energy mode is concentrated in the region
of smaller plasma frequency.

\section{Conclusion\label{sec:conclusions}}

In summary, we have developed a semi-analytical modal approach to
solve Maxwell's equations in the presence of the linearized hydrodynamic
correction. With this approach, the diffraction from a periodic metallic
layer was calculated. The modal method was shown to provide physical
insight to the calculated absorption spectrum, by detailed inspection
of the modal propagation constants. Longitudinal modes with propagation
constants close to zero, were found to generate absorption dips, while
modes with maximal propagation constants were related to the absorption
peaks. Moreover, we presented a general boundary condition matching
scheme, based on the S-matrix algorithm, that incorporated the ABC
needed to match between local and non-local media. In addition, a
variant of the PWM was formulated and used for the first time in order
to calculate the band diagram dispersion of the longitudinal modes.
These numerical tools might provide a useful framework for the design
of plasmonic circuit devices at the very deep nano-scale \cite{engheta}.\appendix

\section{Derivation of the eigenvalue equation\label{sec:eigenvalue-derivation}}

In this Appendix we outline the procedure of derivation of Eq. \eqref{eq:fmm-final}
from Eq. \eqref{one-all-eqn}. Eq. \eqref{eq:1} - \eqref{eq:3} are
rewritten as:\begin{subequations} \label{revised-1-all}

\begin{eqnarray}
\frac{\partial}{\partial z}E_{x} & = & \frac{\partial}{\partial x}E_{z}+j\omega\mu_{0}H_{y}\label{eq:revised-1-1-1}
\end{eqnarray}

\begin{eqnarray}
\frac{\partial}{\partial z}H_{y} & = & +j\omega\varepsilon_{0}E_{x}-jm^{-1}\omega\hat{\omega}^{-2}N_{0}(x)e^{2}E_{x}-\hat{\omega}^{-2}\beta^{2}\frac{\partial}{\partial x}\nabla\cdot J\label{eq:revised-2-1-1}
\end{eqnarray}

\begin{eqnarray}
\frac{\partial}{\partial z}E_{z} & = & -\frac{\partial}{\partial x}E_{x}-\left(i\omega\varepsilon_{0}\right)^{-1}\nabla\cdot J\label{eq:revised-4-1-1}
\end{eqnarray}
Likewise, Eq. \eqref{eq:4} and \eqref{eq:5} are combined to give:

\begin{eqnarray}
\frac{\partial}{\partial z}\nabla\cdot J & = & \beta^{-2}\hat{\omega}^{2}\left(\frac{\partial}{\partial x}H_{y}+j\omega\varepsilon_{0}E_{z}\right)-j\omega N_{0}(x)e^{2}\beta^{-2}m^{-1}E_{z}\label{eq:revised-6-1-1}
\end{eqnarray}
\end{subequations}where $J_{x}$ and $J_{z}$ are eliminated from
Eq. \eqref{eq:revised-2-1-1} and \eqref{eq:revised-6-1-1} according
to:\begin{subequations}

\begin{equation}
J_{z}=-\frac{\partial}{\partial x}H_{y}-j\omega\varepsilon_{0}E_{z}\label{eq:revised-3-1}
\end{equation}

\begin{equation}
J_{x}=-jm^{-1}\omega\hat{\omega}^{-2}N_{0}(x)e^{2}E_{x}-\hat{\omega}^{-2}\beta^{2}\frac{\partial}{\partial x}\nabla\cdot J\label{eq:revised-5-1}
\end{equation}
\end{subequations}Using the parameters defined in Section \ref{sec:fmm-formulation},
Eq. \eqref{revised-1-all} can be written as:\begin{subequations}

\label{revised-der-all}
\begin{equation}
k_{z}E_{x}=k_{x}E_{z}+\tilde{\mu}_{0}H_{y}\label{eq:revised-der-1-1}
\end{equation}

\begin{equation}
k_{z}H_{y}=\tilde{\varepsilon}_{0}E_{x}-\tilde{\varepsilon}_{0}\hat{k}_{0}^{-2}\tilde{\omega}_{p}^{2}(x)E_{x}-\hat{k}_{0}^{-2}\tilde{\beta}^{2}k_{x}\nabla\cdot J\label{eq:revised-der-2-1}
\end{equation}

\begin{equation}
k_{z}E_{z}=\tilde{\mu}_{0}k_{0}^{-2}\nabla\cdot J-k_{x}E_{x}\label{eq:revised-der-4-1}
\end{equation}

\begin{equation}
k_{z}\nabla\cdot J=\tilde{\beta}^{-2}\hat{k}_{0}^{2}k_{x}H_{y}+\tilde{\varepsilon}_{0}\tilde{\beta}^{-2}\hat{k}_{0}^{2}E_{z}-\tilde{\varepsilon}_{0}\tilde{\omega}_{p}^{2}(x)\tilde{\beta}^{-2}E_{z}\label{eq:revised-der-6-1}
\end{equation}
\end{subequations}Eq. \eqref{revised-der-all} can be written in
matrix form, as two first order coupled differential equations:

\begin{subequations}
\begin{equation}
\left[\begin{array}{c}
\mathbf{E_{x}}\\
\mathbf{\nabla\cdot J}
\end{array}\right]\left[\mathbf{K_{z}}\right]=\left[\begin{array}{cc}
\tilde{\mu}_{0}\mathbf{I} & \mathbf{K_{x}}\\
\tilde{\beta}^{-2}\hat{k}_{0}^{2}\mathbf{K_{x}} & \tilde{\varepsilon}_{0}\tilde{\beta}^{-2}\left(\hat{k}_{0}^{2}\mathbf{I}-\mathbf{\Omega_{p}^{2}}\right)
\end{array}\right]\left[\begin{array}{c}
\mathbf{H_{y}}\\
\mathbf{E_{z}}
\end{array}\right]\label{eq:matrix1}
\end{equation}

\begin{equation}
\left[\begin{array}{c}
\mathbf{H_{y}}\\
\mathbf{E_{z}}
\end{array}\right]\left[\mathbf{K_{z}}\right]=\left[\begin{array}{cc}
\tilde{\varepsilon}_{0}\left(\mathbf{I}-\hat{k}_{0}^{-2}\mathbf{\Omega_{p}^{2}}\right) & -\hat{k}_{0}^{-2}\tilde{\beta}^{2}\mathbf{K_{x}}\\
-\mathbf{K_{x}} & \tilde{\mu}_{0}k_{0}^{-2}\mathbf{I}
\end{array}\right]\left[\begin{array}{c}
\mathbf{E_{x}}\\
\mathbf{\nabla\cdot J}
\end{array}\right]\label{eq:matrix2}
\end{equation}
\end{subequations}By eliminating $\mathbf{H_{y}}$ and $\mathbf{E_{z}}$
, Eq. \eqref{eq:matrix1} and \eqref{eq:matrix2} are combined to
obtain Eq. \eqref{eq:fmm-final}.

\section{S-matrix formulation for non-local periodic layers embedded in a
local environment\label{sec:s-matrix}}

\begin{figure}
\begin{centering}
\includegraphics[scale=0.55]{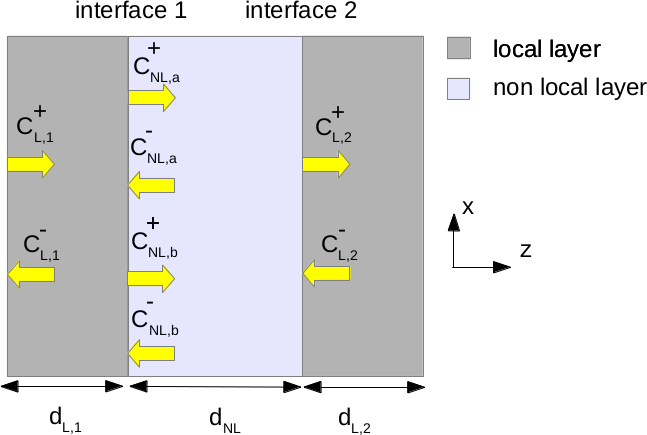}
\par\end{centering}

\caption{Schematic showing a non-local layer embedded in a local environment.
\label{fig:s-matrix-schematic}}
\end{figure}
In this Appendix, the S-matrix for a periodic layer with non-local
response, embedded in a local environment is evaluated. The geometry
is defined in Fig. \ref{fig:s-matrix-schematic}. Two local layers,
labeled as ``$\text{\ensuremath{L_{1}}}$'' and ``$\text{\ensuremath{L_{2}}}$''
surround a non-local slab labeled as ``NL''. The mode amplitude
constants are defined as $C$, with the subscript denoting the layer
index, and the superscript the propagation direction, with ``+''
and ``-'' standing for waves propagating in the positive and negative
z-directions respectively. Each vector of mode amplitudes $C$, has
$N$ elements. The non-local layer therefore supports twice as many
modes than the local layers,  in consistence with the discussion in
Section \ref{sec:fmm-formulation}. For the case that the non-local
layer is non-periodic, the solution of Eq. \eqref{eq:fmm-final} results
in $N$ pure transverse and $N$ pure longitudinal modes. Then, the
mode amplitude constants can be divided into two groups, $C_{\text{NL,a}}$
and $C_{\text{NL,b}}$ with each of these groups associated with either
longitudinal or transverse modes. When the non-local layer is periodic,
no pure longitudinal modes exist in the general case. For consistency,
we keep the division into two groups $C_{\text{NL,a}}$ and $C_{\text{NL,b}}$
so that each mode amplitude vector remains with $N$ elements. However,
the division into these two groups is now arbitrary. To find the mode
amplitude vectors, we employ the S-matrix approach, for which a scattering
matrix $\mathbf{S}$ relates a vector of incident mode amplitudes
\textbf{$\mathbf{a}$ }to a vector of outgoing mode amplitudes $\mathbf{b}$
according to $\mathbf{b}=\mathbf{Sa}$. This approach is considered
as numerically stable in the sense that growing exponential terms
are avoided \cite{lifengli2}. First, the S-matrix for the interfaces
are derived and afterwards they are used to derive the layer S-matrix. 

The two interface S-matrices are defined as:\begin{subequations}\label{interface-all}

\begin{eqnarray}
\left[\begin{array}{ccc}
S_{11}^{(1)} & S_{12}^{(1)} & S_{13}^{(1)}\\
S_{21}^{(1)} & S_{22}^{(1)} & S_{23}^{(1)}\\
S_{31}^{(1)} & S_{32}^{(1)} & S_{33}^{(1)}
\end{array}\right]\left[\begin{array}{c}
P_{\text{L,1}}^{+}C_{\text{L,1}}^{+}\\
C_{\text{NL,a}}^{-}\\
C_{\text{NL,b}}^{-}
\end{array}\right] & = & \left[\begin{array}{c}
C_{\text{NL,a}}^{+}\\
C_{\text{NL,b}}^{+}\\
P_{\text{L,1}}^{-}C_{\text{L,1}}^{-}
\end{array}\right]\label{eq:interface-1}
\end{eqnarray}
\begin{eqnarray}
\left[\begin{array}{ccc}
S_{11}^{(2)} & S_{12}^{(2)} & S_{13}^{(2)}\\
S_{21}^{(2)} & S_{22}^{(2)} & S_{23}^{(2)}\\
S_{31}^{(2)} & S_{32}^{(2)} & S_{33}^{(2)}
\end{array}\right]\left[\begin{array}{c}
C_{\text{L,2}}^{-}\\
P_{\text{NL,a}}^{+}C_{\text{NL,a}}^{+}\\
P_{\text{NL,b}}^{+}C_{\text{NL,b}}^{+}
\end{array}\right] & = & \left[\begin{array}{c}
P_{\text{NL,a}}^{-}C_{\text{NL,a}}^{-}\\
P_{\text{NL,b}}^{-}C_{\text{NL,b}}^{-}\\
C_{\text{L,2}}^{+}
\end{array}\right]\label{eq:interface-2}
\end{eqnarray}
\end{subequations}Here, the phase matrices $P_{\text{L,1}}^{\pm}$
are $N\times N$ diagonal matrices with elements $\exp(jk_{0}k_{z,n}^{(i)\pm}d_{\text{L,1}})$,
with $d_{\text{L,1}}$ the local layer thickness as shown in Fig.
\ref{fig:s-matrix-schematic}. The forward and backward propagating
modes in layer $i$, have propagation constants $k_{z,n}^{(i)+}$
and $k_{z,n}^{(i)-}$ respectively. Since $k_{z,n}^{(i)+}=-k_{z,n}^{(i)-}$,
these phase matrices satisfy $P_{\text{L,i}}^{+}=(P_{\text{L,i}}^{-})^{-1}$.
For the non-local layer the phase matrices are $P_{\text{NL,x}}^{\pm}$
with x=a,b, with the subscripts and superscripts having their obvious
meaning. The phase matrices appear in Eq. \eqref{interface-all} because
the mode amplitudes are defined to have zero phase at the left boundary
of each layer.

Assuming continuity of $E_{x}$, $J_{z}$ and $E_{z}$ we match the
eigenvector matrices of these field quantities at the first interface:

\begin{equation}
\left[\begin{array}{cc}
E_{x,\text{L,1}}^{+} & E_{x,\text{L,1}}^{-}\\
J_{z,\text{L,1}}^{+} & J_{z,\text{L,1}}^{-}\\
E_{z,\text{L,1}}^{+} & E_{z,\text{L,1}}^{-}
\end{array}\right]\left[\begin{array}{c}
P_{\text{L,1}}^{+}C_{\text{L,1}}^{+}\\
P_{\text{L,1}}^{-}C_{\text{L,1}}^{-}
\end{array}\right]=\left[\begin{array}{cccc}
E_{x,\text{NL,a}}^{+} & E_{x,\text{NL,b}}^{+} & E_{x,\text{NL,a}}^{-} & E_{x,\text{NL,b}}^{-}\\
J_{z,\text{NL,a}}^{+} & J_{z,\text{NL,b}}^{+} & J_{z,\text{NL,a}}^{-} & J_{z,\text{NL,b}}^{-}\\
E_{z,\text{NL,a}}^{+} & E_{z,\text{NL,b}}^{+} & E_{z,\text{NL,a}}^{-} & E_{z,\text{NL,b}}^{-}
\end{array}\right]\left[\begin{array}{c}
C_{\text{NL,a}}^{+}\\
C_{\text{NL,b}}^{+}\\
C_{\text{NL,a}}^{-}\\
C_{\text{NL,b}}^{-}
\end{array}\right]\label{eq:interface-matching}
\end{equation}
Where a ``+'' or ``-'' superscript of the eigenvector matrices
represents left and right propagating field quantities respectively.
Rearranging terms in Eq. \eqref{eq:interface-matching} we obtain:

\begin{equation}
\left[\begin{array}{ccc}
E_{x,\text{L,1}}^{+} & -E_{x,\text{NL,a}}^{-} & -E_{x,\text{NL,b}}^{-}\\
J_{z,\text{L,1}}^{+} & -J_{z,\text{NL,a}}^{-} & -J_{z,\text{NL,b}}^{-}\\
E_{z,\text{L,1}}^{+} & -E_{z,\text{NL,a}}^{-} & -E_{z,\text{NL,b}}^{-}
\end{array}\right]\left[\begin{array}{c}
P_{\text{L,1}}^{+}C_{\text{L,1}}^{+}\\
C_{\text{NL,a}}^{-}\\
C_{\text{NL,b}}^{-}
\end{array}\right]=\left[\begin{array}{ccc}
E_{x,\text{NL,a}}^{+} & E_{x,\text{NL,b}}^{+} & -E_{x,\text{L,1}}^{-}\\
J_{z,\text{NL,a}}^{+} & J_{z,\text{NL,b}}^{+} & -J_{z,\text{L,1}}^{-}\\
E_{z,\text{NL,a}}^{+} & E_{z,\text{NL,b}}^{+} & -E_{z,\text{L,1}}^{-}
\end{array}\right]\left[\begin{array}{c}
C_{\text{NL,a}}^{+}\\
C_{\text{NL,b}}^{+}\\
P_{\text{L,1}}^{-}C_{\text{L,1}}^{-}
\end{array}\right]\label{eq:interface-matching-rearranged}
\end{equation}
Comparing Eq. \eqref{eq:interface-matching-rearranged} with \eqref{eq:interface-1},
it is seen that the first interface S-matrix can be expressed by the
eigenvector matrices as:

\begin{subequations}

\begin{equation}
\left[\begin{array}{ccc}
S_{11}^{(1)} & S_{12}^{(1)} & S_{13}^{(1)}\\
S_{21}^{(1)} & S_{22}^{(1)} & S_{23}^{(1)}\\
S_{31}^{(1)} & S_{32}^{(1)} & S_{33}^{(1)}
\end{array}\right]=\left[\begin{array}{ccc}
E_{x,\text{NL,a}}^{+} & E_{x,\text{NL,b}}^{+} & -E_{x,\text{L,1}}^{-}\\
J_{z,\text{NL,a}}^{+} & J_{z,\text{NL,b}}^{+} & -J_{z,\text{L,1}}^{-}\\
E_{z,\text{NL,a}}^{+} & E_{z,\text{NL,b}}^{+} & -E_{z,\text{L,1}}^{-}
\end{array}\right]^{-1}\left[\begin{array}{ccc}
E_{x,\text{L,1}}^{+} & -E_{x,\text{NL,a}}^{-} & -E_{x,\text{NL,b}}^{-}\\
J_{z,\text{L,1}}^{+} & -J_{z,\text{NL,a}}^{-} & -J_{z,\text{NL,b}}^{-}\\
E_{z,\text{L,1}}^{+} & -E_{z,\text{NL,a}}^{-} & -E_{z,\text{NL,b}}^{-}
\end{array}\right]\label{eq:interface-1-final}
\end{equation}
Following a similar procedure, the second interface S-matrix is:

\begin{equation}
\left[\begin{array}{ccc}
S_{11}^{(2)} & S_{12}^{(2)} & S_{13}^{(2)}\\
S_{21}^{(2)} & S_{22}^{(2)} & S_{23}^{(2)}\\
S_{31}^{(2)} & S_{32}^{(2)} & S_{33}^{(2)}
\end{array}\right]=\left[\begin{array}{ccc}
E_{x,\text{NL,a}}^{-} & E_{x,\text{NL,b}}^{-} & -E_{x,\text{L,2}}^{+}\\
J_{z,\text{NL,a}}^{-} & J_{z,\text{NL,b}}^{-} & -J_{z,\text{L,2}}^{+}\\
E_{z,\text{NL,a}}^{-} & E_{z,\text{NL,b}}^{-} & -E_{z,\text{L,2}}^{+}
\end{array}\right]^{-1}\left[\begin{array}{ccc}
E_{x,\text{L,2}}^{-} & -E_{x,\text{NL,a}}^{+} & -E_{x,\text{NL,b}}^{+}\\
J_{z,\text{L,2}}^{-} & -J_{z,\text{NL,a}}^{+} & -J_{z,\text{NL,b}}^{+}\\
E_{z,\text{L,2}}^{-} & -E_{z,\text{NL,a}}^{+} & -E_{z,\text{NL,b}}^{+}
\end{array}\right]\label{eq:interface-2-final}
\end{equation}
\end{subequations}By detaching the phase matrices, Eq. \eqref{interface-all}
can be modified to:\begin{subequations}\label{interface-all-2}

\begin{equation}
\left[\begin{array}{ccc}
I & 0 & 0\\
0 & I & 0\\
0 & 0 & \left(P_{\text{L,1}}^{-}\right)^{-1}
\end{array}\right]\left[\begin{array}{ccc}
S_{11}^{(1)} & S_{12}^{(1)} & S_{13}^{(1)}\\
S_{21}^{(1)} & S_{22}^{(1)} & S_{23}^{(1)}\\
S_{31}^{(1)} & S_{32}^{(1)} & S_{33}^{(1)}
\end{array}\right]\left[\begin{array}{ccc}
P_{\text{L,1}}^{+} & 0 & 0\\
0 & I & 0\\
0 & 0 & I
\end{array}\right]\left[\begin{array}{c}
C_{\text{L,1}}^{+}\\
C_{\text{NL,a}}^{-}\\
C_{\text{NL,b}}^{-}
\end{array}\right]=\left[\begin{array}{c}
C_{\text{NL,a}}^{+}\\
C_{\text{NL,b}}^{+}\\
C_{\text{L,1}}^{-}
\end{array}\right]
\end{equation}

\begin{equation}
\left[\begin{array}{ccc}
\left(P_{\text{NL,a}}^{-}\right)^{-1} & 0 & 0\\
0 & \left(P_{\text{NL,b}}^{-}\right)^{-1} & 0\\
0 & 0 & I
\end{array}\right]\left[\begin{array}{ccc}
S_{11}^{(2)} & S_{12}^{(2)} & S_{13}^{(2)}\\
S_{21}^{(2)} & S_{22}^{(2)} & S_{23}^{(2)}\\
S_{31}^{(2)} & S_{32}^{(2)} & S_{33}^{(2)}
\end{array}\right]\left[\begin{array}{ccc}
I & 0 & 0\\
0 & P_{\text{NL,a}}^{+} & 0\\
0 & 0 & P_{\text{NL,b}}^{+}
\end{array}\right]\left[\begin{array}{c}
C_{\text{L,2}}^{-}\\
C_{\text{NL,a}}^{+}\\
C_{\text{NL,b}}^{+}
\end{array}\right]=\left[\begin{array}{c}
C_{\text{NL,a}}^{-}\\
C_{\text{NL,b}}^{-}\\
C_{\text{L,2}}^{+}
\end{array}\right]
\end{equation}
\end{subequations}Preforming the matrix multiplications in Eq. \eqref{interface-all-2}
and rearranging terms, we obtain two homogeneous equations:\begin{subequations}\label{interface-all-3}

\begin{equation}
\left[\begin{array}{cccccc}
S_{11}^{(1)}P_{\text{L,1}}^{+} & S_{12}^{(1)} & S_{13}^{(1)} & -I & 0 & 0\\
S_{21}^{(1)}P_{\text{L,1}}^{+} & S_{22}^{(1)} & S_{23}^{(1)} & 0 & -I & 0\\
P_{\text{L,1}}^{+}S_{31}^{(1)}P_{\text{L,1}}^{+} & P_{\text{L,1}}^{+}S_{32}^{(1)} & P_{\text{L,1}}^{+}S_{33}^{(1)} & 0 & 0 & -I
\end{array}\right]\left[\begin{array}{c}
C_{\text{L,1}}^{+}\\
C_{\text{NL,a}}^{+}\\
C_{\text{NL,b}}^{+}\\
C_{\text{NL,a}}^{-}\\
C_{\text{NL,b}}^{-}\\
C_{\text{L,1}}^{-}
\end{array}\right]=0
\end{equation}

\begin{equation}
\left[\begin{array}{cccccc}
P_{\text{NL,a}}^{+}S_{11}^{(2)} & P_{\text{NL,a}}^{+}S_{12}^{(2)}P_{\text{NL,a}}^{+} & P_{\text{NL,a}}^{+}S_{13}^{(2)}P_{\text{NL,b}}^{+} & -I & 0 & 0\\
P_{\text{NL,b}}^{+}S_{21}^{(2)} & P_{\text{NL,b}}^{+}S_{22}^{(2)}P_{\text{NL,a}}^{+} & P_{\text{NL,b}}^{+}S_{23}^{(2)}P_{\text{NL,b}}^{+} & 0 & -I & 0\\
S_{31}^{(2)} & S_{32}^{(2)}P_{\text{NL,a}}^{+} & S_{33}^{(2)}P_{\text{NL,b}}^{+} & 0 & 0 & -I
\end{array}\right]\left[\begin{array}{c}
C_{\text{L,2}}^{-}\\
C_{\text{NL,a}}^{+}\\
C_{\text{NL,b}}^{+}\\
C_{\text{NL,a}}^{-}\\
C_{\text{NL,b}}^{-}\\
C_{\text{L,2}}^{+}
\end{array}\right]=0
\end{equation}
\end{subequations}Combining the two equations in \eqref{interface-all-3},
we obtain a system of equations that connect the incident amplitudes
($C_{\text{L,1}}^{+}\text{ and }C_{\text{L,2}}^{-}$) with the outgoing
($C_{\text{L,2}}^{+}\text{ and }C_{\text{L,1}}^{-}$) and internal
amplitudes ($C_{\text{NL,a}}^{+},C_{\text{NL,b}}^{+},C_{\text{NL,a}}^{-}\text{ and }C_{\text{NL,b}}^{-}$)
:\begin{subequations}\label{s-matrix-final-1}

\begin{eqnarray}
S_{\text{layer}}\left[\begin{array}{c}
C_{\text{L,1}}^{+}\\
C_{\text{L,2}}^{-}
\end{array}\right] & = & \left[\begin{array}{c}
C_{\text{L,2}}^{+}\\
C_{\text{L,1}}^{-}\\
C_{\text{NL,a}}^{+}\\
C_{\text{NL,b}}^{+}\\
C_{\text{NL,a}}^{-}\\
C_{\text{NL,b}}^{-}
\end{array}\right]
\end{eqnarray}
\begin{eqnarray}
S_{\text{layer}} & \equiv & A^{-1}B
\end{eqnarray}

\begin{eqnarray}
A & \equiv & \left[\begin{array}{cccccc}
0 & 0 & I & 0 & -S_{12}^{(1)} & -S_{13}^{(1)}\\
0 & 0 & 0 & I & -S_{22}^{(1)} & -S_{23}^{(1)}\\
0 & I & 0 & 0 & -P_{\text{L,1}}^{+}S_{32}^{(1)} & -P_{\text{L,1}}^{+}S_{33}^{(1)}\\
0 & 0 & -P_{\text{NL,a}}^{+}S_{12}^{(2)}P_{\text{NL,a}}^{+} & -P_{\text{NL,a}}^{+}S_{13}^{(2)}P_{\text{NL,b}}^{+} & I & 0\\
0 & 0 & -P_{\text{NL,b}}^{+}S_{22}^{(2)}P_{\text{NL,a}}^{+} & -P_{\text{NL,b}}^{+}S_{23}^{(2)}P_{\text{NL,b}}^{+} & 0 & I\\
I & 0 & -S_{32}^{(2)}P_{\text{NL,a}}^{+} & -S_{33}^{(2)}P_{\text{NL,b}}^{+} & 0 & 0
\end{array}\right]
\end{eqnarray}
 
\begin{eqnarray}
B & \equiv & \left[\begin{array}{cc}
S_{11}^{(1)}P_{\text{L,1}}^{+} & 0\\
S_{21}^{(1)}P_{\text{L,1}}^{+} & 0\\
P_{\text{L,1}}^{+}S_{31}^{(1)}P_{\text{L,1}}^{+} & 0\\
0 & P_{\text{NL,a}}^{+}S_{11}^{(2)}\\
0 & P_{\text{NL,b}}^{+}S_{21}^{(2)}\\
0 & S_{31}^{(2)}
\end{array}\right]
\end{eqnarray}
\end{subequations}$S_{\text{layer}}$ is a $6N\times2N$ matrix,
which can be divided into two submatrices according to:\begin{subequations}\label{s-matrix-final-2}

\begin{eqnarray}
S_{\text{layer}} & \equiv & \left[\begin{array}{c}
S_{\text{external}}\\
S_{\text{internal}}
\end{array}\right]
\end{eqnarray}

\begin{eqnarray}
S_{\text{external}} & \equiv & \left[\begin{array}{cc}
S_{\text{layer},11} & S_{\text{layer},12}\\
S_{\text{layer},21} & S_{\text{layer},22}
\end{array}\right]
\end{eqnarray}

\begin{eqnarray}
S_{\text{internal}} & \equiv & \left[\begin{array}{cc}
S_{\text{layer},31} & S_{\text{layer},32}\\
S_{\text{layer},41} & S_{\text{layer},42}\\
S_{\text{layer},51} & S_{\text{layer},52}\\
S_{\text{layer},61} & S_{\text{layer},62}
\end{array}\right]
\end{eqnarray}
\end{subequations}Eq. \eqref{s-matrix-final-1} and \eqref{s-matrix-final-2}
provide a complete description of all mode amplitudes. $S_{\text{external}}$
can be regarded as the ordinary $2N\times2N$ S-matrix, which couples
transverse only modes in the first local layer to transverse only
modes in the second local layer. From $S_{\text{internal}}$ the internal
(i.e. non-local) amplitudes $C_{\text{NL,a}}^{+},C_{\text{NL,b}}^{+},C_{\text{NL,a}}^{-}\text{ and }C_{\text{NL,b}}^{-}$
can be obtained.
\begin{acknowledgments}
This research was supported by the AFOSR. A. Y. acknowledges the support
of the CAMBR and Brojde fellowships.\end{acknowledgments}

\end{document}